\documentclass[pra,aps,twocolumn,superscriptaddress,a4paper,showpacs]{revtex4}
\usepackage{graphicx,amssymb,amsmath}

\newcommand{\ket}[2][]{#1\lvert #2 #1\rangle}

\newcommand{\down}{\ensuremath{\downarrow}}
\newcommand{\up}{\ensuremath{\uparrow}}
\newcommand{\dg}{^\dagger}
\newcommand{\PT}{T_A}
\newcommand{\nn}{\nonumber \\}

\begin{document}

\title{Two qubits can be entangled in two distinct temperature regions}

\author{Dominic W. Berry}
\affiliation{School of Physical Sciences, The
University of Queensland, Queensland 4072, Australia}
\author{Mark R. Dowling}
\email{dowling@physics.uq.edu.au} \affiliation{School of Physical Sciences, The
University of Queensland, Queensland 4072, Australia}

\date{\today}

\begin{abstract}
We have found that for a wide range of two-qubit Hamiltonians the
canonical-ensemble thermal state is entangled in two distinct temperature
regions. In most cases the ground state is entangled; however we have also found
an example where the ground state is separable and there are still two regions.
This demonstrates that the qualitative behavior of entanglement with temperature
can be much more complicated than might otherwise have been expected; it is not
simply determined by the entanglement of the ground state, even for the simple
case of two qubits. Furthermore, we prove a finite bound on the number of
possible entangled regions for two qubits, thus showing that arbitrarily many
transitions from entanglement to separability are not possible. We also provide
an elementary proof that the spectrum of the thermal state at a lower
temperature majorizes that at a higher temperature, for any Hamiltonian, and use
this result to show that only one entangled region is possible for the special
case of Hamiltonians without magnetic fields.
\end{abstract}

\pacs{03.67.Mn, 03.65.Ud}

\maketitle

\section{Introduction}
A topic that has emerged recently within the field of quantum information is the
study of entanglement in spin systems (see~\cite{nielsenphd,arnesen2001} for
early examples).  Entanglement is often considered necessarily a low-temperature
phenomenon that becomes less important as the temperature is increased.  It was
therefore surprising that in~\cite{nielsenphd} an example for two qubits was
given where the ground state ($T=0$) is separable but the thermal state is
entangled at higher temperatures. This behavior can be understood as due to the
presence of low-lying excited states that are entangled.  Thus at least two
qualitatively-different entanglement scenarios are possible for two qubits
(apart from the uninteresting case of no entanglement at any temperature); (i)
the ground state is entangled, and hence the thermal state is entangled at low
temperatures up to a critical temperature, $T_S$, above which it is separable,
and (ii) the ground state is separable, but the thermal state is entangled for
temperatures within some finite range $(T_1,T_S)$ (and separable again above
$T_S$).

The question we address here is what other entanglement scenarios are possible
for two qubits. We were stimulated to ask this question by~\cite{osenda2005}
which studies the generic behavior of thermal entanglement as a function of
temperature. There it was shown that the generic behavior is closed intervals in
temperature where the thermal state is separable interspersed with open
intervals of entanglement. Examples of the two types mentioned above were given
for two qubits.  Also two examples of qubit-qutrit systems ($d_1=2$ and $d_2=3$,
where $d_j$, $j=1,2$ are the dimensions of the two subsystems) were given where
there are two distinct entangled regions; in one case the ground state is
entangled, and in the other case the ground state is separable.

Non-monotonic behavior of thermal entanglement has also been observed for qubit
spin chains~\cite{rosciPRL04,amico2006}. Ref.\ \cite{rosciPRL04} studies the
reduced state of two nearby qubits in particular spin chains and finds examples
of two distinct entangled regions in temperature. Ref.\ \cite{amico2006} uses a
multipartite entanglement measure, and observes three regions. In
\cite{cavalcanti2005} transitions from separability to entanglement are studied
as a type of phase transition using geometric arguments about the set of
separable states.

Here we restrict to the case of just two qubits (as opposed to the reduced state
of two qubits out of a large chain), and present a class of Hamiltonians for
which most have a value of the magnetic field strength such that the ground
state is entangled and there are two entangled regions.  In addition, we present
an example of a Hamiltonian outside this class that has a separable ground state
and, again, two entangled regions. Thus we find that all the classes of behavior
for the thermal states of qubit-qutrit systems found in Ref.\ \cite{osenda2005}
are also observed for the two-qubit case.

These results raise the question of how many distinct ``entangled regions'' in
temperature are possible. Our numerical search failed to find Hamiltonians with
more entangled regions for two qubits, indicating that this is the most
complicated behavior. We show that for a class of commonly considered
Hamiltonians --- those without magnetic fields --- it is impossible to obtain
more than one entangled region. In addition we derive upper bounds on the number
of entangled regions in the general case.

This paper is set out as follows. In Sec.~\ref{sec:gen} we present results for
the dimer case of the spin-chain Hamiltonian studied in Ref.~\cite{rosci2004},
then give our general class of two-qubit Hamiltonians. In Sec.~\ref{sec:cou} we
give an example of a Hamiltonian in our class that does not exhibit two
entangled regions, and show that small perturbations are sufficient to give two
entangled regions. In Sec.~\ref{sec:unen} we give our example with a separable
ground state and two entangled regions. We derive bounds on the total number of
entangled regions possible for two qubits in Sec.~\ref{sec:upper}, specialize to
the case with zero magnetic field in Sec.~\ref{sec:belldiag}, and then conclude
in Sec.~\ref{sec:conc}.

\section{General Case}
\label{sec:gen}

The two-qubit case of the $XYX$ spin-chain studied in Ref.~\cite{rosciPRL04} is
of the following form:
\begin{equation}
\label{eq:rosci} H=-J\left[X_1X_2-Z_1Z_2+h(Z_1+Z_2) \right].
\end{equation}
where $\{ X_j, Y_j, Z_j \}$ are the Pauli sigma matrices acting on qubits
$j=1,2$, $J$ is a coupling constant with dimensions of energy and $h$ is a
dimensionless parameter corresponding to the magnetic fields experienced by the
qubits. The results for this Hamiltonian on two-qubits were given in the inset
of Fig.\ 4 in an early version of this paper~\cite{rosci2004}. This figure
showed that two entangled regions were obtained, though this aspect of the
results was not discussed.

An isolated system (i.e.\ not exchanging particles with the environment) in
thermal equilibrium with a bath at temperature $T$ will reach the
canonical-ensemble thermal state $\rho(T)$ given by
\begin{equation}
\rho(T) = \frac{e^{-\beta H}}{\mathcal{Z}},
\end{equation}
where $\beta=1/k_BT$ and $\mathcal{Z} = \mathrm{Tr}[\exp(-\beta H)]$ is the
partition function.

\begin{figure}
\begin{centering}
\includegraphics[width=8cm]{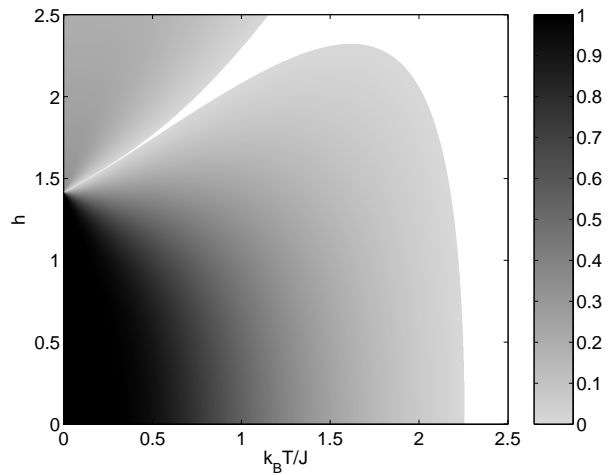}
\end{centering}
\caption{Concurrence as a function of temperature and $h$ for two qubits coupled
according to~(\ref{eq:rosci}).} \label{fig:rosciplot}
\end{figure}

In Fig.~\ref{fig:rosciplot} we plot the concurrence~\cite{wootters1998} as a
function of (scaled) temperature and the scaled magnetic field $h$ for the
canonical-ensemble thermal state~\footnote{The concurrence is nonzero if and
only if a two-qubit (possibly mixed) state is entangled.}. We see that there is
a significant range of values of $h$ (approximately $\sqrt{2} < h < 2.36$) such
that the entanglement as a function of temperature behaves as claimed --- there
are two entangled regions, one at low temperatures, $[0,T_1)$, and then another
distinct region at higher temperatures, $(T_2,T_S)$.

A general two-qubit Hamiltonian may be written in the form (omitting any global
phase shift)
\begin{equation}
H=\vec X_1^T R \vec X_2 + A \vec X_1 + B \vec X_2,
\end{equation}
where $R$ is a real $3\times 3$ matrix, $A$ and $B$ are real row vectors, and
$\vec X_k=[X_k ~Y_k ~Z_k]^T$, $k\in\{1,2\}$. It is possible to add local unitary
operations before and after the Hamiltonian without affecting the entanglement of
the thermal state. This is because
\begin{equation}
e^{-\beta V_1\otimes V_2 H V_1\dg\otimes V_2\dg}= V_1\otimes V_2e^{-\beta H}
V_1\dg\otimes V_2\dg,
\end{equation}
where $V_1$ and $V_2$ are local unitary operations on subsystems 1 and 2. These
local unitaries act to rotate the $\vec X_k$ vectors, giving $V_k \vec X_k
V_k\dg = O_k \vec X_k$ for some orthogonal matrices $O_k$. Thus
\begin{align}
H' &= V_1\otimes V_2 H V_1\dg\otimes V_2\dg \nn &=\vec X_1^T O_1^T R O_2\vec X_2
+ A O_1\vec X_1 + B O_2\vec X_2.
\end{align}
If the local operations are chosen such that $O_1$ and $O_2$ are the orthogonal
matrices which result from a singular-value decomposition of $R$, then $O_1^T R
O_2$ is a positive diagonal matrix \cite{andrew}.

Hence $H'$ is of the form
\begin{align}
\label{eq:cgen} H'&=J[\alpha_xX_1X_2+\alpha_yY_1Y_2+\alpha_zZ_1Z_2
+h(\beta_{x1}X_1 \nn & +\beta_{x2}X_2+\beta_{y1}Y_1+\beta_{y2}Y_2+
\beta_{z1}Z_1+\beta_{z2}Z_2)],
\end{align}
where the $\alpha_j$ are positive, and the $\beta_{jk}$ are real. Note that the
local unitaries do not remove the local component of the Hamiltonian. If it were
possible to use different local unitaries before and after the Hamiltonian, the
local component of the Hamiltonian could be removed entirely \cite{dur2001}.
However, this would change the entanglement of the thermal state.

Hamiltonians of the form \eqref{eq:cgen} are the most general two-qubit
Hamiltonians for the problem of thermal entanglement. Now we introduce a class
of Hamiltonians which is slightly restricted, in that we require $\beta_{j1}
=\beta_{j2}=\beta_j$. This is equivalent to requiring the magnetic field to be
homogeneous. These Hamiltonians may be written as
\begin{align}
\label{eq:gen} H&=J\left\{\alpha_xX_1X_2+\alpha_yY_1Y_2+\alpha_zZ_1Z_2 \right.
\nn & \left. +h[\beta_x(X_1+X_2)+\beta_y(Y_1+Y_2)+\beta_z(Z_1+Z_2)]\right\}.
\end{align}
As before, the $\alpha_j$ are positive, and the $\beta_j$ are real. To determine
properties of these Hamiltonians, random Hamiltonians were generated, and for
each it was determined if there exists a value of $h$ such that there are two
entangled regions. The $\alpha_j$ were chosen at random in the interval $[0,1)$,
and the $\beta_j$ at random in the interval $(-1,1)$. From a sample of $2^{20}$
of these Hamiltonians, it was found that all had a value of $h$ such that there
are two entangled regions.

Arbitrary two-qubit Hamiltonians were also tested. These were generated
according to the Gaussian unitary ensemble. Each Hamiltonian was divided into a
local part $H_L$ and a nonlocal part $H_N$, and Hamiltonians of the form
$H_h=H_N+hH_L$ were tested. It was found that, out of $2^{12}$ samples, there
were 106 such that there was a value of $h$ for which $H_h$ has two entangled
regions. This gives the overall probability for this behavior for arbitrary
Hamiltonians as $2.59\pm 0.25\%$.

\section{Examples}
\label{sec:cou} Although we have shown that it is extremely common for
Hamiltonians of the form \eqref{eq:gen} to have two entangled regions, not all
exhibit this behavior. For example, consider the $XY$ interaction with a
magnetic field, as studied by Wang \cite{wang2001}:
\begin{align}
\label{eq:wang} H&=J\left\{X_1X_2+Y_1Y_2+h(Z_1+Z_2)]\right\}.
\end{align}
Wang found that it was possible for the thermal entanglement to be zero for
$T=0$ but nonzero for $T>0$. Wang also considered the anisotropic $XY$
interaction, but without a magnetic field. In neither case were two regions
found.

\begin{figure}
\begin{centering}
\includegraphics[width=8cm]{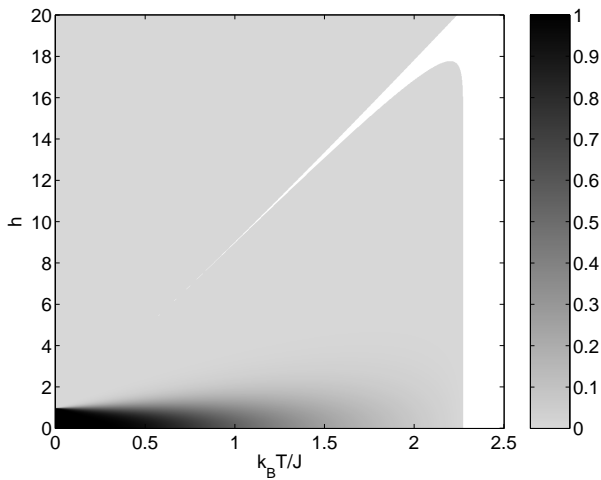}
\end{centering}
\caption{Concurrence as a function of temperature and $h$ for two qubits coupled
according to~(\ref{eq:hamiltonian}) with $\gamma = 10^{-6}$.}
\label{fig:thermalplot2}
\end{figure}

It turns out that we can vary the Hamiltonians very slightly from this example,
and again recover the two entangled regions. For example, consider the
anisotropic $XY$ interaction
\begin{equation}
\label{eq:hamiltonian} H = J\left[ (1+\gamma) X_1 X_2 + (1-\gamma) Y_1 Y_2 +
h(Z_1 + Z_2) \right].
\end{equation}
For $\gamma=0$ this is the Hamiltonian of Eq.\ \eqref{eq:wang}. However, for
$\gamma$ equal to just $10^{-6}$, we again recover the two entangled regions
(see Fig.\ \ref{fig:thermalplot2}).

\begin{figure}
\begin{centering}
\includegraphics[width=8cm]{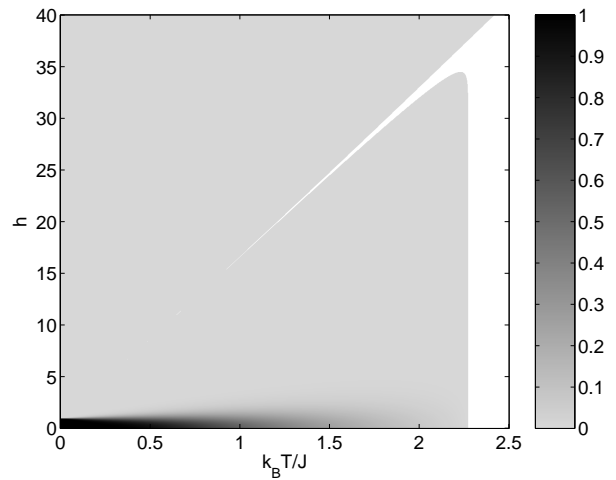}
\end{centering}
\caption{Concurrence as a function of temperature and $h$ for two qubits coupled
according to~(\ref{eq:pert}) with $\delta=10^{-6}$.} \label{fig:misal}
\end{figure}

Another perturbation which recovers the two entangled regions is that where the
magnetic field is not exactly aligned on the $z$-axis:
\begin{equation}
\label{eq:pert} H=J\left\{X_1X_2+Y_1Y_2+h[Z_1+Z_2+\delta(X_1+X_2)]\right\}.
\end{equation}
For $\delta\ll 1$ this is the $XY$ interaction with a misaligned transverse
magnetic field. The concurrence for $\delta=10^{-6}$ is shown in Fig.\
\ref{fig:misal}. Even with this very small misalignment in the magnetic field,
the distinct entangled regions are again seen.

\section{Separable ground state}
\label{sec:unen}

The next most complicated case is that where the ground state is separable, so
the thermal state is separable at $T=0$, but there are still two entangled
regions. As local unitaries do not alter the entanglement, one can arbitrarily
choose the separable ground state without loss of generality. Thus to
numerically search for such examples we took the ground state to be
$|00\rangle$. The other eigenstates and eigenenergies were then chosen at
random.

\begin{figure}
\begin{centering}
\includegraphics[width=8cm]{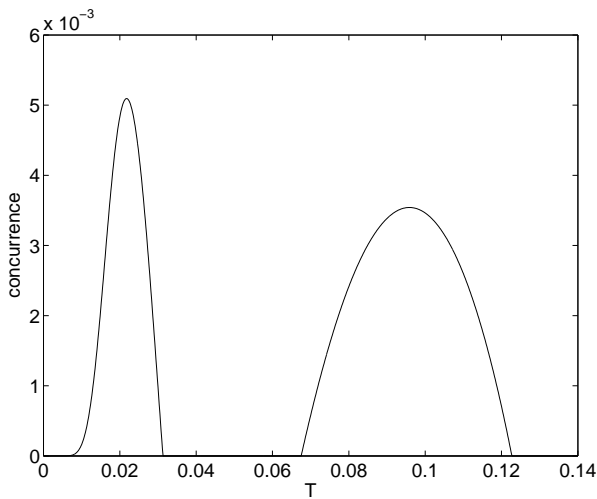}
\end{centering}
\caption{Concurrence as a function of temperature for two-qubit thermal state
under the Hamiltonian \eqref{eq:unen}.} \label{fig:unen}
\end{figure}

The example found was (after rounding the coefficients)
\begin{align}
\label{eq:unen} H &= 0.006(X_1X_2+Y_1Y_2)+0.03(X_1Y_2-Y_1X_2) \nn
&+0.02(Z_1X_2-X_2)+(Z_1Y_2-Y_2)/10 \nn &+(X_1Z_2-X_1)/14+Z_1Z_2/7-Z_1/4-Z_2/5.
\end{align}
The concurrence as a function of temperature is as shown in Fig.\
\ref{fig:unen}. There are two distinct regions of entanglement, with a separable
ground state. Note that there appears to be a finite region without entanglement
for low temperature. However, for much of this region the concurrence is
extremely small (less than $10^{-20}$) but nonzero. This indicates that the
thermal state may be completely separable only for zero temperature, and the
entanglement for small temperatures is not observed due to finite precision.

\section{Bound on number of entangled regions}
\label{sec:upper}

We used numerical techniques to search for examples of more complicated
scenarios from the hierarchy, i.e.\ three or more entangled regions, but were
unable to find any. Of course no numerical technique can be exhaustive so it
remains an intriguing possibility that even more entangled regions are possible,
even for two qubits. We now show that there is, in fact, a finite upper bound,
17, on the number of entangled regions for two qubits. However it remains
entirely plausible that this bound is not tight and the above examples of two
entangled regions represent the most complicated behavior possible for two
qubits.

To analytically bound the number of entangled regions for two qubits we use the
well-known fact that a two-qubit mixed state is entangled or separable depending
on whether its partial transpose with respect to one of the qubits has a
negative eigenvalue or not. Therefore, by solving
\begin{equation}
\det[\rho(T)^{\PT}]=0,
\end{equation}
where $\PT$ denotes the partial transpose with respect to the first subsystem,
we find the transitions between entangled and separable thermal states. We may
scale the energies so that the minimum energy eigenvalue is zero. If we multiply
by $\mathcal{Z}$, the resulting equation is polynomial in $e^{-1/k_BT}$ with
noninteger powers and $35$ terms~\footnote{The polynomial contains terms of the
form $e^{-E/k_BT}$, where $E$ is the sum of any four, possibly nondistinct,
energy eigenvalues; $35$ is the number of distinct terms of this form.}.

Provided the ratios of the energy eigenvalues are all rational, the polynomial
has integer powers in $x=e^{-r/k_BT}$ for some constant $r$. To place a bound on
the number of solutions, we first take the derivative of the polynomial, then
apply Descartes' rule of signs. The derivative of the polynomial has no more
than 35 terms, and so has no more than 34 sign changes. By Descartes' rule there
are no more than $34$ (positive) zeros of the derivative, and no more than $34$
turning points of the polynomial.

It is easily seen that, provided the derivative has no more than 34 zeros, there
are no more than $17$ regions where the polynomial is negative. In the two-qubit
case, the partial transpose has no more than one negative
eigenvalue~\cite{sanpera1998}; thus the determinant is negative if and only if
the partial transpose is negative.  Hence there can be no more than $17$
entangled regions. Thus we obtain a finite limit on the number of entangled
regions, though this is much larger than the number of intervals which have been
found numerically. In practice the number of sign changes is likely to be far
less than $34$, although we do not see a way of showing this analytically.

In the case where there are irrational ratios of the energy eigenvalues, the
situation is more complicated because the polynomial has noninteger powers. In
this case, we can achieve an arbitrarily close approximation of the Hamiltonian
with rational energies. In the case where a function $f$ is the limit of a
sequence of functions $f_n$, it is not possible for $f$ to have more turning
points than $f_n$. The only situation where $f'$ (where the prime indicates the
derivative) can have more zeros than $f'_n$ is when $f'_n$ has an extremum which
approaches zero in the limit $n\to\infty$, and is only exactly equal to zero for
$f'$. However, this zero would correspond to a point of inflection, rather than
an extremum, for $f$. Thus we find that the polynomial in $e^{-1/k_BT}$ can have
no more than 17 regions where it is negative, and the limit on the number of
entangled regions must hold for irrational powers also.

For a qubit coupled to a qutrit entangled mixed states must still have a
non-positive partial transpose~\cite{horodecki2001}, however there is a
complication due to the fact that the partial transpose can have more than one
negative eigenvalue. The main problem in this case is that, at a point where
$\det[\rho(T)^{\PT}]$ changes sign, the state could be separable for
$\det[\rho(T)^{\PT}]=0$, but entangled for slightly higher or lower
temperatures. This could happen if one of the eigenvalues passes from positive
to negative, while another passes from negative to zero to negative.

However, despite this possibility it can be seen that the number of turning
points of $\det[\rho(T)^{\PT}]$ still provides an upper bound on the number of
entangled regions. For a qubit-qutrit system there are six energy levels and so
up to $462$ terms in the polynomial corresponding to $\det[\rho(T)^{\PT}]$
(assuming rational eigenvalues). There are therefore no more than $461$ turning
points, even in the limit of irrational eigenvalues. Combined with the fact that
the state must be separable at high temperature, this implies that the number of
entangled regions can be no higher than $462$ (some of them may be separated by
single points in temperature where the system is separable).

More generally, for the case of two subsystems of arbitrary dimensions, $d_1$
and $d_2$, one might hope to put a finite upper bound on the number of
entangled regions as a function of $d_1$ and $d_2$.  However, in higher
dimensions entangled mixed states do not necessarily have a non-positive partial
transpose. In fact it has recently been shown that even the problem of
distinguishing separable and entangled mixed states is $NP$-hard in arbitrary
dimension~\cite{gurvits2003}. It is therefore unlikely that this approach will
yield upper bounds for higher dimensional systems.

\section{Zero magnetic field}
\label{sec:belldiag}

Although we were unable to definitively answer the question of what entanglement
scenarios are possible for an arbitrary two-qubit Hamiltonian, we can for a
certain class of Hamiltonian --- those that have no local terms (corresponding
to a magnetic field), and only interaction terms. A Hamiltonian without local
terms may be written in the form
\begin{equation}
H=\vec X_1^T R \vec X_2,
\end{equation}
As in Sec.\ \ref{sec:gen}, we can apply local unitaries without altering the
entanglement of the thermal state. These simplify the Hamiltonian to a form that
is diagonal in the Bell Basis.
\begin{equation}
H' = V_1\otimes V_2 H V_1\dg\otimes
V_2\dg=J(\alpha_xX_1X_2+\alpha_yY_1Y_2+\alpha_zZ_1Z_2).
\end{equation}
The Bell basis is a set of maximally entangled states
\begin{align}
\ket{\phi^{\pm}}&=\frac{1}{\sqrt{2}}\left(\ket{\up} \ket{\up} \pm \ket{\down} \ket{\down}\right)\,, \\
\ket{\psi^{\pm}}&=\frac{1}{\sqrt{2}}\left(\ket{\up} \ket{\down} \pm \ket{\down} \ket{\up}\right)\,,
\end{align}
where $Z \ket{\up} = \ket{\up}$, $Z \ket{\down} = -\ket{\down}$ (up and down
spins if the qubit is a spin 1/2 quantum system).

To determine the behavior of the entanglement, we compare the state at two
different temperatures, $T_1$ and $T_2$, such that $T_1<T_2$. We first note that
the eigenvalues for $\rho(T_2)$ are majorized by those for $\rho(T_1)$. It is
straightforward to show this result for all bipartite thermal states (not just
those for two-qubit systems). For $T_1<T_2$, $\beta_1>\beta_2$. Therefore, for
$\Delta E\ge(\le)\, 0$, we have $e^{-\beta_2\Delta E}\ge(\le)\,
e^{-\beta_1\Delta E}$. Taking the energy eigenvalues $E_i$ to be sorted into
nondescending order, we have
\begin{align}
\sum_{i=1}^k e^{-\beta_1(E_i-E_k)} &\ge \sum_{i=1}^k e^{-\beta_2(E_i-E_k)}, \\
\sum_{i=k+1}^d e^{-\beta_2(E_i-E_k)} &\ge \sum_{i=k+1}^d e^{-\beta_1(E_i-E_k)}.
\end{align}
Multiplying gives
\begin{align}
\sum_{i=1}^k e^{-\beta_1E_i}\sum_{i=k+1}^d e^{-\beta_2E_i} &\ge
 \sum_{i=1}^k e^{-\beta_2E_i}\sum_{i=k+1}^d e^{-\beta_1E_i}, \\
\sum_{i=1}^k e^{-\beta_1E_i}\sum_{i=1}^d e^{-\beta_2E_i} &\ge
 \sum_{i=1}^k e^{-\beta_2E_i}\sum_{i=1}^d e^{-\beta_1E_i}.
\end{align}
Hence
\begin{equation}
\sum_{i=1}^k e^{-\beta_1E_i}/\mathcal{Z}_1 \ge \sum_{i=1}^k
e^{-\beta_2E_i}/\mathcal{Z}_2,
\end{equation}
which is the result claimed.

Now, for density operators $\rho_1$ and $\rho_2$ such that the eigenvalues for
$\rho_2$ are majorized by those for $\rho_1$, we have \cite{nielsen1999}
\begin{equation}
\rho_2 = \sum_j p_j U_j\dg \rho_1 U_j,
\end{equation}
where the unitaries $U_j$ permute the eigenstates~\footnote{In the general case
the $U_j$ would also need to include a possible change in the eigenstates from
$\rho_1$ to $\rho_2$. We do not need that here, because the eigenstates are
unchanged.}.

For the specific case where the Hamiltonian is diagonal in the Bell basis, the
eigenstates are just the Bell basis states. In this case, if the state $\rho_1$
is separable, then all states $U_j\dg \rho_1 U_j$ obtained by permuting the
eigenstates are also separable. To show this, we first note that it is not
necessary to preserve phase when permuting the eigenstates, because any phase
cancels out in the density matrix. In order to permute the eigenstates (without
regard for phase), it is sufficient to show that it is possible to perform three
swaps between eigenstates. All permutations may be constructed from these three
swaps. We may obtain three swaps between Bell basis pairs using local unitaries
as follows:
\begin{align}
\ket{\phi^+}\leftrightarrow\ket{\phi^-}&:e^{i\pi Z/4}\otimes
e^{i\pi Z/4}, \\
\ket{\psi^+}\leftrightarrow\ket{\psi^-}&:e^{-i\pi Z/4}\otimes
e^{i\pi Z/4}, \\
\ket{\phi^-}\leftrightarrow\ket{\psi^+}&:\mathcal{H}\otimes \mathcal{H},
\end{align}
where $\mathcal{H} = (X+Z)/\sqrt{2}$ is the Hadamard operation.
Thus we find that it is possible to perform any permutation of the Bell basis
using local operations, so if $\rho_1$ is separable, each of the states $U_j\dg
\rho_1 U_j$ is separable. Hence the state $\rho_2$ must be separable.

Thus, for Hamiltonians that are diagonal in the Bell basis, the eigenvalues of
$\rho(T_2)$ for $T_2>T_1$ are majorized by those for $\rho(T_1)$, so if
$\rho(T_1)$ is separable, then so is $\rho(T_2)$. Thus we can not have a
situation where the thermal entanglement increases with temperature. As we may
simplify any two-qubit Hamiltonian without local terms to a form which is
diagonal in the Bell basis, this result holds for all two-qubit Hamiltonians
without local terms.

\section{Conclusions}
\label{sec:conc}

We have presented a class of Hamiltonians for which almost all examples have a
value of the magnetic field such that the thermal state has two distinct
entangled regions. One example of this class previously appeared as a figure in
an online paper~\cite{rosci2004}, but this aspect of the results was not
discussed explicitly. This result is somewhat surprising as one may have
expected that the small Hilbert space for two qubits would mean that only one
entangled region were possible.

There are, however, particular cases from this class where distinct regions do
not occur, for example the isotropic $XY$ interaction with a transverse magnetic
field \cite{wang2001}. However, we find that if the interaction is perturbed
only slightly by making it anisotropic or misaligning the magnetic field,
distinct regions do occur. This suggests that those cases where the distinct
regions do not occur are a set of measure zero in this class.

It is also possible for there to be two entangled regions when the ground state
is separable. In contrast to the case where there are two regions and an
entangled ground state, this behavior is extremely rare. It was necessary to
test millions of Hamiltonians before an example of this form was found.

We have also shown that certain features of the examples are necessary in order
to observe the distinct entangled regions.  We proved that for Hamiltonians
without local terms (i.e. no magnetic field) the entanglement must necessarily
decrease with increasing temperature, so only one entangled region is possible
(at low temperatures).

For general two-qubit Hamiltonians we showed, by considering zeros of the
determinant of the partial transpose of the thermal density matrix, that there
can be no more than $17$ entangled regions. Thus arbitrarily many transitions
from entanglement to separability are not possible for two qubits (or a qubit
and a qutrit).

\begin{acknowledgments}
This project has been supported by the Australian Research Council. We thank
Andrew Doherty, Michael Nielsen, David Poulin, and other members of the QiSci
problem solving group at the University of Queensland for valuable discussions.
\end{acknowledgments}

\end{document}